\documentclass[twocolumn,superscriptaddress,nobalancelastpage,aps,pra]{revtex4-2}
\usepackage[latin1]{inputenc}
\usepackage{graphicx}
\usepackage{amsmath}
\usepackage{amsthm}
\usepackage{amssymb}
\usepackage{latexsym}
\usepackage{array}
\usepackage{float}
\usepackage{amsfonts}
\usepackage{mathrsfs}
\usepackage{verbatim}
\usepackage{dsfont}
\usepackage{xcolor}
\usepackage[font=small,labelsep=space]{caption}
\usepackage{subcaption}

\begin{document}

\title{Detection of quantum phase transition in spin-1 chain through multipartite high-order correlations}

\author{Dongkeun Lee}
\affiliation{Department of Physics, Sogang University, 35, Baekbeom-ro, Mapo-gu, Seoul, 04107, Republic of Korea}
\affiliation{Research Institute for Basic Science, Sogang University, 35, Baekbeom-ro, Mapo-gu, Seoul, 04107, Republic of Korea}

\author{Adel Sohbi}
\affiliation{School of Computational Sciences, Korea Institute for Advanced Study, Seoul, 02455, Republic of Korea}

\author{Wonmin Son}
\email{sonwm@physics.org}
\affiliation{Department of Physics, Sogang University, 35, Baekbeom-ro, Mapo-gu, Seoul, 04107, Republic of Korea}

\date{\today}

\begin{abstract}
We design a Bell inequality that is violated by correlations obtained from the ground states of XXZ spin-1 chain with on-site anisotropies at the region of phase transition. In order to detect such correlations in spin-1 systems we exploit the formalism of generalized Bell inequality via the use of multipartite and high-order correlations. We observe sharp violation in the vicinity of quantum phase transition between the so-called large-$D$ and AFM phase. Interestingly, the violation of our Bell inequality is manifested by the change of the XXZ spin-1 chain ground state to a Greenberger-Horne-Zeilinger (GHZ)-like state at the critical region. Our results provide the first characterization of quantum phase transition via the violation of Bell-type constraint by correlations in the XXZ spin-1 chain with multi-body correlations and high-order measurements.
\end{abstract}

\maketitle


\section{Introduction}

A set of quantum correlations in a composite quantum system can be used to demonstrate non-locality when it exceeds the bound imposed by a local hidden-variable model \cite{Bell:1965, Brunner:2014}. Non-locality is a feature of quantum systems that has no counterpart in any classical systems. Since the first experimental confirmation of this property \cite{Aspect:1982}, experimental observations have been challenged rigorously but with the recent advancement of experiment apparatus, known loopholes have been closed \cite{Hensen:2015, Giustina:2015, Shalm:2015}. As its practical use, it is known that non-locality is a crucial resource for device-independent (DI) quantum information processing to achieve security going beyond the classical limit {\it e.g.}\cite{Acin:2007,Pironio:2009,Pironio:2010}. 

Recently, investigations on the quantum non-locality have been extended to various many-body systems \cite{Tura:2014,Schmied:2016,Engelsen:2017,Wagner:2017} following the studies of many-body entanglement a decades ago \cite{Amico:2008,DeChiara:2018}. These recent studies have revealed the importance of quantum non-locality in many-body systems and reported that Bell-type correlation is possible to witness many-body quantum criticality \cite{Piga:2019}.
In particular, Mermin-type multipartite correlation \cite{Mermin:1990}, has been used to test many spin systems \cite{Batle:2010,Dai:2017,Bao:2020,Niezgoda:2020}. Alternatively, the ground-state energy has also been employed to identify non-locality in various many-body systems \cite{Tura:2017}. Similarly, it was demonstrated that the ground state of the Ising model with infinite-range interactions in the external field can produce non-local correlations at finite temperature \cite{Fadel:2018}. The most recent investigations for Bell inequalities involving the linear combination of two-body correlations were extended in various ways \cite{Frerot:2021, Muller:2021}. Our knowledges about the characterization of non-locality in many-body systems are either limited in their analysis by two-level systems or at most the two-body correlations as the behavior of the quantum state was observed within the limit of the first-order correlations only.

Multiparttite Bell-type correlations in the local composite $d$-dimensional systems are not yet widely studied in many-body systems. This is mainly because for the complete characterization of such systems Bell-type correlations are required to include multi-body and high-order correlations which can be numerically challenging to do in practice \cite{Brunner:2014}. On the other hand, by exploiting symmetric properties of such systems, it is possible to be make the correlation analysis simpler. Thus, we exploit a formalism that was derived under the generic scenario of Bell inequalities for multipartite and high-dimensional systems \cite{Bae:2018}. Using this formalism, we inspect the non-locality in the one-dimensional spin-1 model and depict it in phase diagram. As a result of our investigation, non-locality is found in the region of large exchanges and strong on-site anisotropies along the line of quantum phase transition. 
The violation provides the first quantification of quantum criticality using the generalized Bell correlation of high-order for the spin-1 system. 

\section{Spin-1 XXZ chain with on-site anisotropy}

We consider the spin-1 XXZ chain model with on-site anisotropy which is known to have various non-trivial phases \cite{Haldane:1983,Schulz:1986,Chen:2003}. We are interested in whether such phases could be characterized through Bell-type non-local correlations. The formal description of the model is
\begin{align}\label{eq:spin1H}
\hat{H}=\sum^N_{l=1} \hat{S}^x_l\hat{S}^x_{l+1}+\hat{S}^y_l\hat{S}^y_{l+1}+J_z\hat{S}^z_l\hat{S}^z_{l+1}
+ D \sum^N_{l=1}(\hat{S}^z_l)^2,
\end{align}
where $\hat{S}^{a}_l$ with $a\in\{x,y,z\}$ denotes the spin-1 operator for $l$-th site and $\hat{S}^{a}_{N+1}=\hat{S}^{a}_{1}$ for the periodic boundary conditions. The parameters $J_z$ and $D$ signifies the coupling strength of  the two-body spin-$z$ interaction and the anisotropy in the $z$ direction, respectively.

With an appropriate adjustment of the exchange anisotropy $J_z$ and the onsite anisotropy $D$, the ground state $|\psi_{gs}\rangle$ can undergo various phases, one of which is the Haldane phase, renowned for the symmetry-protected topological order \cite{Haldane:1983,Schulz:1986,Chen:2003}. In this investigation, we focus on the region of non-negative parameters where we found the violation of the generalized Bell-type inequality \cite{Bae:2018}. When $J_z\ge 0$ and $D\ge 0$, the spin-1 chain in Eq.\eqref{eq:spin1H} possesses three different phases and they are the Haldane, the anti-ferromagnetic (AFM), and the large-$D$ phases for the different values of $J_z$ and $D$ \cite{Chen:2003}. The ground state of this model is able to be obtained by using the exact diagonalization as well as the density-matrix renormalization group (DMRG) method for a small-scale chain (small $N$) \cite{tenpy}.

\section{Generalized Bell inequality for multipartite system}
\subsection{Bell Correlation}
For the first, let us consider a generic Bell scenario with $N$ parties that share a many-particle state performing $k$-local measurements on each particle. Each party exploits the $k$ possible choices of measurements that yield $d$ different outcomes. The scenario for the generalized Bell-type inequality is suggested under general symmetries \cite{Bae:2018}. This means that the set of Bell-type correlations that appears in this inequality takes into account two symmetries for the equal distribution of measurements and the site permutation.

For simplicity, we adopt the case of two different measurements for each party only. From \cite{Bae:2018}, the generalized Bell correlation for the scenario $(N,d)$ is given by the expectation value of the Bell operator $\mathcal{B}=\langle\psi|\hat{\mathcal{B}}|\psi\rangle$ and the Bell operator with two local measurements $\hat{A}$ and $\hat{B}$ reads
\begin{align}\label{eq:n2dBellop}
\hat{\mathcal{B}} = \sum^{d-1}_{n=1} \left[ f_n\,\bigotimes^{N}_{l=1} \left(\hat{A}^{c_l n}_{l}+\omega^{c_l\frac{n}{2}}\hat{B}^{c_l n}_{l}\right)\right] + h.c.,
\end{align}
where $\omega=\exp(2\pi i/d)$ and \textit{h.c.} denotes the Hermitian conjugate. The arbitrary parameter $c_l$ for a party $l$ takes a value of either $+1$ or $-1$ and the value $-1$ of $c_l$ on an operator implies its conjugate transpose, i.e. $\hat{O}^{-1} =\hat{O}^\dagger$. The weight $f_n$, a complex number for an integer $n$, determines the types of Bell inequality. The measurement operators $\hat{A}_l$ and $\hat{B}_l$ are set to have the eigenvalues $\omega^n$ for the $n$-th outcome.
Here, from the adequate choices of measurement basis of $\hat{A}_l$ and $\hat{B}_l$ 
\footnote{The measurement operators are choosen as $\hat{A}_l=\sum_{\alpha=0}^{d-1}\omega^{\alpha}|\alpha\rangle\langle\alpha|, \hat{B}=\sum_{\beta=0}^{d-1}\omega^{\beta}|\beta\rangle\langle\beta|$, where the measurement bases are obtained as $|\alpha\rangle=\frac{1}{\sqrt{d}}\sum_{z=0}^{d-1}\omega^{z(\alpha-\nu_l)}|z\rangle$ and $|\beta\rangle=\frac{1}{\sqrt{d}}\sum_{z=0}^{d-1}\omega^{z(\beta+1/2-\nu_l)}|z\rangle$ from a computational bases $|z\rangle$},
we obtain
\begin{equation}
\omega^{n\nu_l}\hat{J}^n_l\equiv[\hat{A}^n_l + \omega^{\frac{n}{2}} \hat{B}^n_l]/2,~~\mbox{where}~~
\hat{J}_l\equiv\sum_{z=1}^{d-1}|z-1\rangle\langle z|
\end{equation}
is a lower shift matrix that is similar to the angular momentum lowering operator. Furthermore, in order to satisfy the particular symmetry that will be explained in the next section, we specify the constant $c_l$ to take the value $1$ for odd $l$ and $-1$ for even $l$.


\subsection{Local realistic bound} If a set of correlations is allowed by the local hidden variable model, it is possible to establish a Bell-type inequality $\mathcal{B}\le \beta_{LR}$ that has a real valued upper bound, $\beta_{LR}$, often called the classical bound \cite{Brunner:2014}. Through Fourier analysis, it is possible to show that the Bell-type correlation $\mathcal{B}$ is described as a convex sum of the joint probabilities of relevant measurement outcomes \cite{Lee:2007}.
If the set of probabilities is represented in a vector space, then the local realistic bound $\beta_{LR}$ can be obtained from the convex properties of probability and from the Farkas lemma \cite{Bae:2018}. 

Since we are considering spin-1 systems, it is enough to consider the local realistic bound of the $(N,3)$ scenario of Eq.\eqref{eq:n2dBellop}, which can be derived as
\begin{align} \label{eq:lrb}
\beta_{LR} = &\max_{\{\alpha_l\}} \left[2\sum_{\{m_l\}} |f_1|\cos \Theta_1 + |f_2|\cos \Theta_2 \right],
\end{align}
where 
\begin{align}
\Theta_n = \theta_n + \frac{2\pi n}{3}\vec{c}\cdot\vec{\alpha}+ \frac{\pi n }{3}\vec{c}\cdot\vec{m}
\end{align}
and $f_n = |f_n|e^{i\theta_n}$ for $1\le l \le N$ \cite{Bae:2018}. $\vec{m}=(m_1,\cdots,m_N)$ signifies the particular choice of measurements for all parties,
$\vec{\alpha}=(\alpha_1(m_1),\cdots,\alpha_N(m_N))$ indicates the outcome configuration for a measurement choice $\vec{m}$,
and $\vec{c}=(c_1,\cdots, c_N)$ with $c_l\in\{1,-1\}$.

In order to compute the local realistic bound $\beta_{LR}$, we are generally confronted with the strategy of optimized term-counting for all the deterministic vectors $\vec{\alpha}$, i.e., all the possible outcomes with respect to the possible choices of measurements. It is notable that Eq.(\ref{eq:lrb}) is conventionally solved via a linear programming instance. However, such instance becomes hard to solve as the problem size scales exponentially with the number of systems \cite{Brunner:2014}.
Detailed numerical methods we here use are addressed in the Appendix~\ref{append:B}.
\section{Results}\label{sec:res}
\subsection{Quantum violation at the phase transition}\label{sec:qv}
We analyze the violation of Bell inequalities with $N$-body correlations through the $(N,3)$-class Bell-type inequalities $\mathcal{B}\le\beta_{LR}$ with the ground state of the one-dimensional spin-1 chain in Eq.\eqref{eq:spin1H}. By taking advantage of the different symmetries that leave the Hamiltonian invariant, the $N$-body correlations of the ground states of the Hamiltonian described in Eq.\eqref{eq:n2dBellop} can be rewritten in a much compact form. 
The Hamiltonian of the spin-1 XXZ chain with on-site anisotropy in Eq.\eqref{eq:spin1H} is invariant under the translational symmetry and conserves the magnetization along the $z$ axis. 
This implies a nonzero value of the $N$-body correlation $\big<\bigotimes_{l=1}^{N}\hat{J}^{c_ln}_l\big>$ under the the constraint  $\sum_l c_l =0$ for even $N$, which is obvious in the context of the spin-1 basis \footnote{In this paper, we make use of the notation of the local basis $\{|\beta\rangle| \beta=0,1,2\}$ but the spin-1 basis $\{|+\rangle,|0\rangle,|-\rangle\}$ is also mainly used in spin systems. From these notations, $\sqrt{2}\hat{J}_l = \hat{S}^+_l$ for the spin-1 particle.}.
Moreover, any ground state with real coefficients satisfies $\langle \hat{J}^n_1\hat{J}^{-n}_2 \cdots \rangle = \langle \hat{J}^{-n}_1\hat{J}^n_2 \cdots\rangle$. 
In accordance with these properties, the correlation can take the following simplified form 
\begin{align}\label{eq:n23Bellcorr}
\mathcal{B} = 2^{N+1} \sum_{n=1}^2 |f_n|\cos\left(\theta_n-n\theta_\nu\right) \Bigg<\bigotimes^{N}_{l=1}\hat{J}^{c_ln}_{l}\Bigg>,
\end{align}
where $\theta_\nu = 2\pi\nu_{tot}/3$ and $\nu_{tot}=\sum_l c_l\nu_l$ is the accumulation of local phase shift. In what follows, we only consider spin-1 systems of a small size chain, $i.e.$ $N=4$, $6$, $8,$ and $10$.

Since the local realistic bound given in Eq.\eqref{eq:lrb} is determined by the weights $f_1$ and $f_2$, we take the ratio of Bell correlation to local realistic bound $\mathcal{B}/\beta_{LR}$ as a function of $\tilde{f}\equiv|f_2|/|f_1|$, $\theta_1$, $\theta_2$, and $\theta_\nu$. We numerically maximize the correlation $\mathcal{B}$ over these variables (see the Appendix~\ref{append:C}). We find that the angles $\theta_{1,\max}=(-1)^{N/2}\pi/2$, $\theta_{2,\max}=\pi$, and $\theta_{\nu,\max}=\pi/2$ result in the maximum value of $\mathcal{B}/\beta_{LR}$. The sign of $\theta_{1,\max}$ is flipped by the number of particles due to the negative value of the first-order correlation $\langle\hat{J}_1\hat{J}^{-1}_2\cdots\rangle$ when $N/2$ is odd but the alternating sign does not affect the value of the classical bound.

\begin{figure}[b!]
    \begin{subfigure}[]{0.37\textwidth}
        \includegraphics[width=\textwidth]{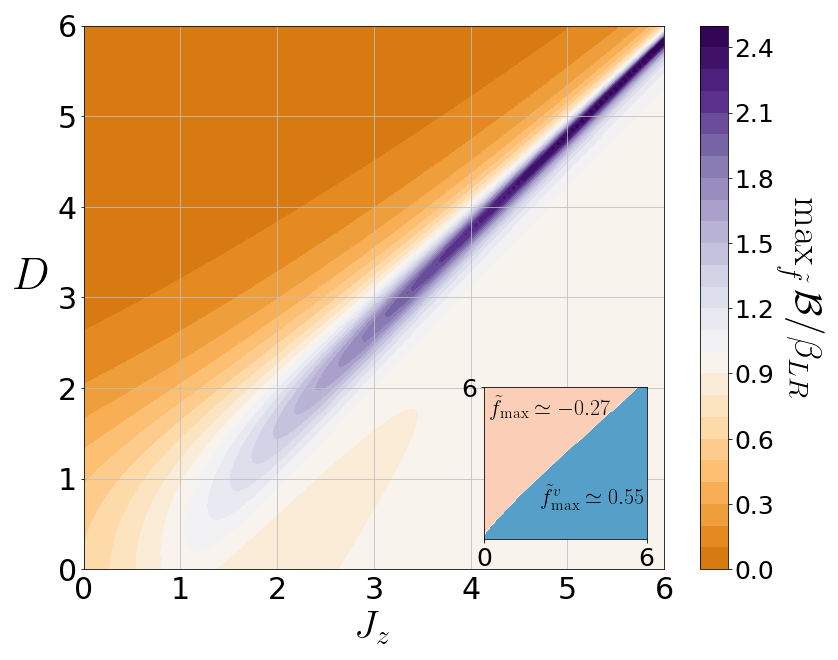}
    \end{subfigure}
    \caption[]{In the $J_z$-$D$ plane the ratio of Bell correlation to the local realistic bound $\mathcal{B}/\beta_{LR}$ maximized over $\tilde{f}$ for $N=8$. Inset : weights $\tilde{f}_{max}$ choices in specific region. The violation of a Bell inequality $\mathcal{B}\ge\beta_{LR}$ arises (in purple). This violation signifies the transition of quantum phase from the large D phase to AFM phase.
}
    \label{fig:NLN8}
\end{figure}

After substituing, $\theta_{1,\max}$, $\theta_{2,\max}$ and $\theta_{\nu,\max}$, in Eq.\eqref{eq:n23Bellcorr} the optimization of $\mathcal{B}/\beta_{LR}$ can be done only over $\tilde{f}$.
We find two regions with two different weights $\tilde{f}$ that give the maximal value of $\mathcal{B}/\beta_{LR}$ for varying $J_z$ and $D$. In Fig.~\ref{fig:NLN8}, is plotted the maximum value of $\mathcal{B}/\beta_{LR}$ for $N=8$ in the $J_z$-$D$ plane. The two regions have been depicted in the inset of Fig.~\ref{fig:NLN8}.

\begin{figure}[b!]
    \begin{subfigure}[]{0.23\textwidth}
        \caption{$J_z=6$}
        \includegraphics[width=\textwidth]{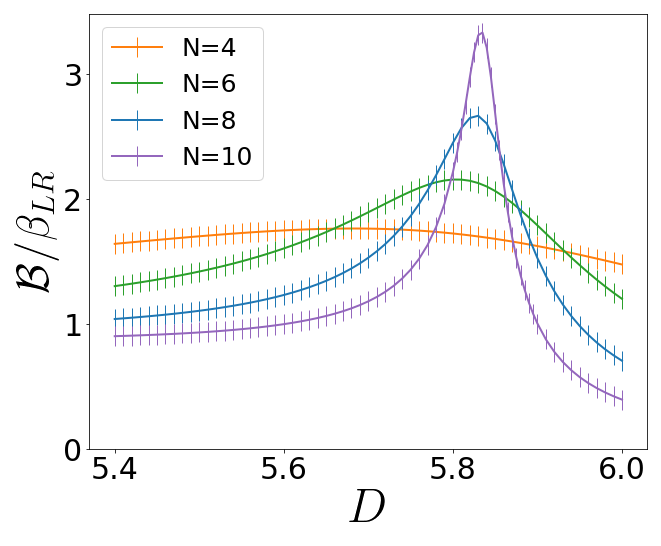}
    \end{subfigure}
    \begin{subfigure}[]{0.23\textwidth}
        \caption{$J_z=12$}
        \includegraphics[width=\textwidth]{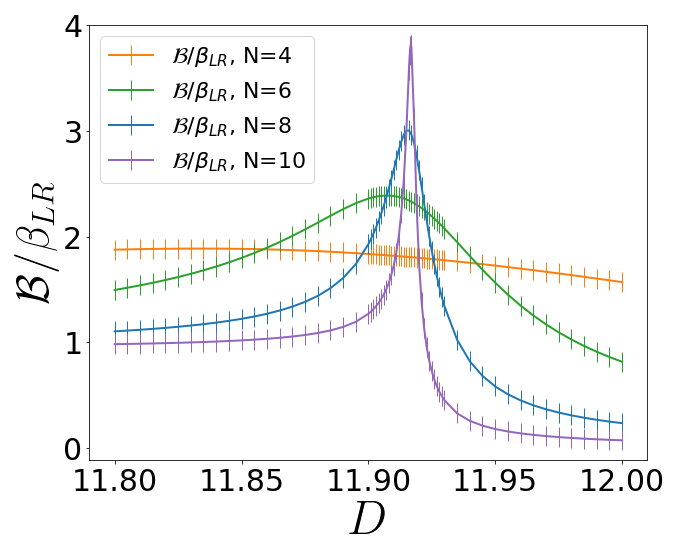}
    \end{subfigure}
    \caption[]{Value of $\mathcal{B}/\beta_{LR}$ with $f^v_{\max}$ for $J_z=\{6,12\}$. The maximum value of $\mathcal{B}/\beta_{LR}$ occurs at the first-order phase transition for large $J_z$. At this criticality, $\mathcal{B}/\beta_{LR}$ increases as $N$ grows large and behaves as a exponential function of $N$, $\mathcal{B}/\beta_{LR}\sim \gamma^N$.}
    \label{fig:NL_Jz}
\end{figure}

Bell nonlocality is then revealed where the ground state gives the violation of a Bell inequality $\mathcal{B}/\beta_{LR}\le 1$. It is notable that the violation of a generalized Bell inequality $\mathcal{B}/\beta_{LR}$ with $\tilde{f}^v_{\max}$ only appears in the vicinity of the criticality between the large-$D$ and AFM phases. This criticality is known as the first-order phase transition where the discontinuity of the staggered magnetization in large-scale systems appears. As $J_z$ and $D$ increase the violation of this Bell inequality increases and occurs more clearly in Fig.~\ref{fig:NLN8}.
The value of $\tilde{f}^v_{\max}$ depends on the number of spins $N$ and in the Table~\ref{table} gives a summary of all the values of $\tilde{f}^{v}_{\max}$ we found.
In Fig.~\ref{fig:NL_Jz}, we show that the region where we observe a violation is located between the large-$D$ and AFM phases for different values of $N$. Moreover, the violation reaches its maximal value along the critical line.

\begingroup
\begin{table}[t]
\begin{tabular}{c | c | c | c }
\hline\hline
$N$ & $\tilde{f}^v_{\max}$ & $b$ in $|\psi_{\max}\rangle$ & $\langle\psi_{\max}|\hat{\mathcal{B}}|\psi_{\max}\rangle$ / $\beta_{LR}$ \\
\hline
$4$ &    1.039 &  0.5798 & 1.950   \\
$6$ &  0.7423 & -0.5599 & 2.470   \\
$8$ &  0.5502 & 0.5457 & 3.119   \\
$10$ & 0.4114 & -0.5348 & 3.973   \\
\hline\hline
\end{tabular}
\caption{Numerical results for the weight $\tilde{f}^v_{\max}$ by solving $\max_{\tilde{f}}\mathcal{B}/\beta_{LR}$ in the ground state near the criticality.
For a given $\tilde{f}^v_{\max}$, the coefficient $b$ of the state $|\psi_{\max}\rangle$ and the maximal value of the Bell correlation $\langle\psi_{\max}|\hat{\mathcal{B}}|\psi_{\max}\rangle$ divided by $\beta_{LR}$ are obtained by solving $\max_{|\psi\rangle} \langle\psi|\hat{\mathcal{B}}|\psi\rangle$.
}
\label{table}
\end{table}
\endgroup

For fixed measurements one can identify a state that can give the maximal value of $\mathcal{B}$ described in Eq.\eqref{eq:n23Bellcorr}. In order to characterize the source of violation obtained by the ground state of the XXZ spin-1 chain, we first identify such state for our Bell inequality and then compare it the ground state during the quantum phase transion between the large-$D$ and AFM phases.
For the Bell inequality with fixed weight $\tilde{f}^v_{\max}$, we denote the quantum state which gives the maximum of Eq.\eqref{eq:n23Bellcorr} by
\begin{equation}\label{eq:psi_max}
|\psi_{\max}\rangle=b|02\cdots\rangle+\sqrt{1-2b^2}|11\cdots\rangle+b|20\cdots\rangle,
\end{equation}
where a real $b$ is determined by $\tilde{f}^v_{\max}$ given in Table~\ref{table}.
This state is the eigenvector with the highest eigenvalue of $\hat{\mathcal{B}}$ using the weight $\tilde{f}^v_{\max}$ where the Bell operator is $\hat{\mathcal{B}}=\sum^2_{n=1}(-1)^{nN/2}|f_n|\hat{J}^n_1\cdots\hat{J}^{-n}_N + h.c.$ with $|f_1|=1$ and $|f_2|=\tilde{f}_{\max}^v$.
Note that for $\tilde{f}=1$, the state $|\psi_{\max}\rangle$ is the generalized GHZ state.

\begin{figure}[b]
    \centering
    \includegraphics[width=0.37\textwidth]{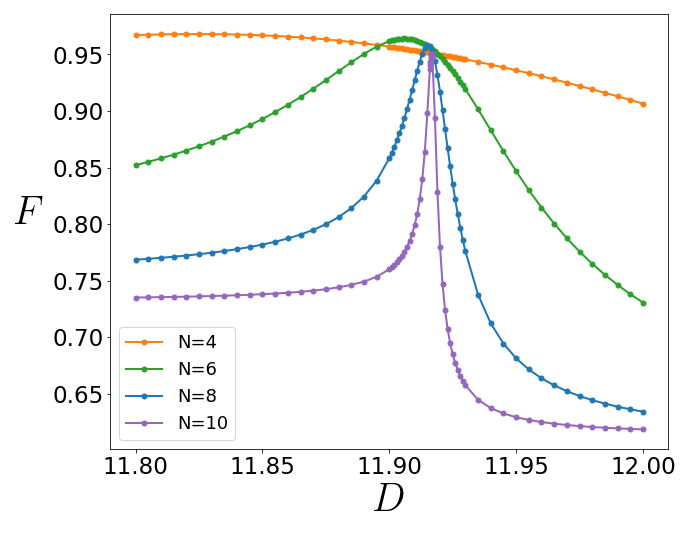}
    \caption{The fidelity $F$ between the ground state and the state $|\psi_{\max}\rangle$ defined in Eq.\eqref{eq:psi_max} for $J_z=12$ near the first-order phase transitions up to the particle number $N\le 10$.
    We can see that the maximum of the fidelity $F$ occurs at the criticality.}
    \label{fig:fidelity}
\end{figure}

In order to explain the source of violation we compare the ground state during the quantum phase transition between the large-$D$ and the AFM phases to the state that maximally violate our Bell inequality given in Eq.\eqref{eq:psi_max}.
For that purpose, we comppute the quantity $F\equiv|\langle\psi_{\max} |\psi_{gs}\rangle|$ which is the fidelity between the ground state $|\psi_{gs}\rangle$ and the state $|\psi_{\max}\rangle$ defined in Eq.\eqref{eq:psi_max}. In Fig.~\ref{fig:fidelity},  the fidelity is shown for $N\in\{4,6,8,10\}$ and $J_z=12$ when $D$ varies between $11.80$ and $12$.
The different points correspond to the evaluation of the fidelity after the computation of the ground state for each value of $J_z$ and $D$.
This region is characterized by a quantum phase transition \cite{Chen:2003} and the maximum violation of our Bell inequality (see Fig.~\ref{fig:NL_Jz}(b)).
Interestingly, we see that the fidelity is higher when the ground state is evaluated closer to the criticality.
For instance, for $N=10$, we can see in Fig.~\ref{fig:fidelity} for $J_z=12$ and when the value of $D$ increases from $D\approx 11.80$ (in the AFM phase) we have $F\approx 0.74$ and increases to its maximum value $F\approx0.95$ at the criticality to finally go down to $F\approx 0.62$ in the large-$D$ phase.
This means that, during this quantum phase transition, the ground state becomes closer to the state that give the maximum violation of our Bell inequality.
This process is then reflected on the violation as we see that the fidelity in Fig.~\ref{fig:fidelity} and the violation in Fig.~\ref{fig:NL_Jz}(b) follows the same tendency.

\begin{figure}[t!]
    \begin{subfigure}[]{0.23\textwidth}
        \caption{$J_z=6$}
        \includegraphics[width=\textwidth]{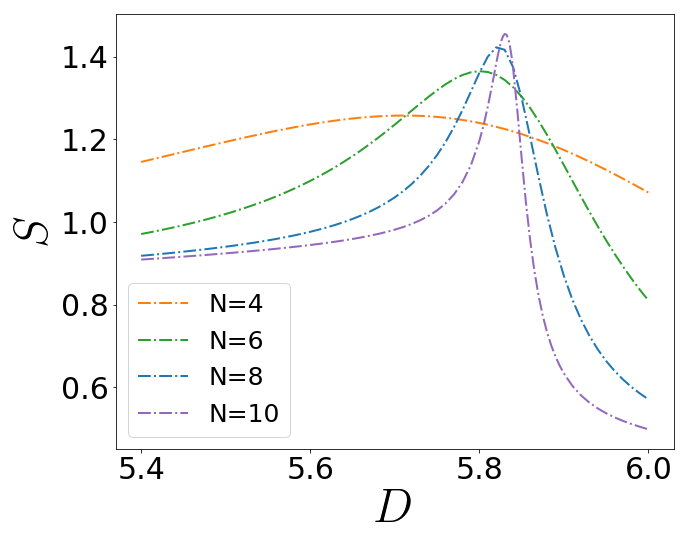}
    \end{subfigure}
    \begin{subfigure}[]{0.23\textwidth}
        \caption{$J_z=12$}
        \includegraphics[width=\textwidth]{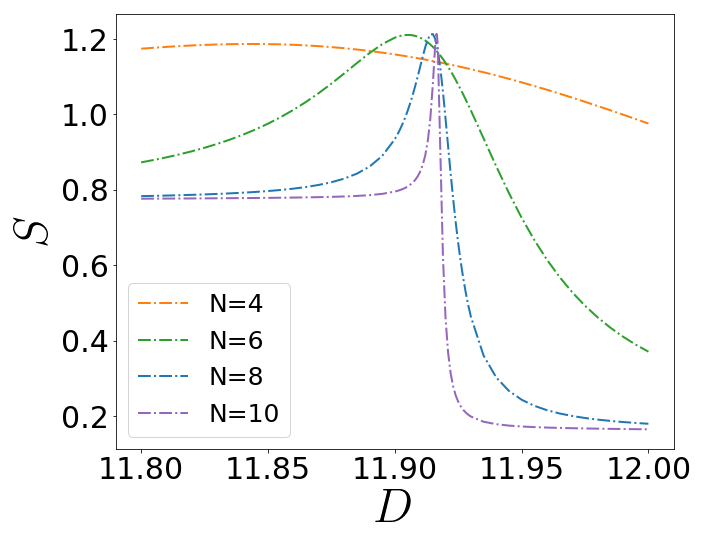}
    \end{subfigure}
    \caption[]{Entanglement entropy $S(\hat{\rho}_{N/2})$ in the vicinity of the first-order phase transition for exchange anisotropy $J_z=6$ and $J_z=12$.
    }
    \label{fig:EE_Jz}
\end{figure}

For large positive $J_z$ and $D$, the ground state undergoes two different phases, the AFM and large-$D$ phases. In our numerical analysis, we identify that the ground state is $|11\cdots 1\rangle$ in the large-$D$ phase and $|02\cdots 02\rangle$ + $|20\cdots 20\rangle$ in the AFM phase, which has been already characterized in \cite{Chen:2003}. We also characterize entanglement in the ground state by computing the bipartite entanglement, $S(\hat{\rho}_{N/2})$, between two halves of the spin system during the quantum phase transition. In Fig.~\ref{fig:EE_Jz}, we show the value of the bipartite entropy for $N\in\{4,6,8,10\}$ and $J_z=\{6,12\}$ at the quantum phase transion. The bipartite entanglement follows exactly the same trend as the violation presented in Fig.~\ref{fig:NL_Jz}(a) ($J_z=6$) and Fig.~\ref{fig:NL_Jz}(b) ($J_z=12$) and the fidelity in Fig.~\ref{fig:fidelity} ($J_z=12$).

\subsection{Scaling behavior of Bell correlation divided by local realistic bound}
The various properties of the spin model Eq.\eqref{eq:spin1H} arising for small $J_z$ and $D$ has been characterized in \cite{Chen:2003} and in the region of positive parameters, three types of the phase exist: the Haldane, the large-$D$, and the AFM phases and quantum phase transitions emerge between them. In our investigation, we did not find violation of the Bell inequalities $\mathcal{B}\le\beta_{LR}$ based on Eq.\eqref{eq:n2dBellop} and Eq.\eqref{eq:lrb} near $J_z=D=0$.
A noticable difference is that the entanglement entropy $S(\hat{\rho}_{N/2})$ is dominant for small $J_z$ and $D$ where the nonviolation of $\mathcal{B}\le\beta_{LR}$ is observed in Fig.~\ref{fig:NLN8}. We thus focus on the first-order phase transition between the large-$D$ and AFM phases for analysis of the scaling behavior.

It is known that the entanglement entropy manifests a logarithmic behavior $S(\hat{\rho}_{N/2})\sim\log N$ in the gapless systems while a saturation for the size of the system $N$ in the gapped systems \cite{Vidal:2003}.
Since the large-$D$-to-AFM phase transitions is the first order, the entanglement entropy reaches saturation for finite-size $N$.
Whereas, in Fig.~\ref{fig:NL_Jz}, the ratio $\mathcal{B}/\beta_{LR}$ with $\tilde{f}^v_{\max}$ becomes higher and narrower as the number of the spin-1 particles increases.
We demonstrate that scaling behavior $\mathcal{B}/\beta_{LR}\sim \gamma^N$ with a real $\gamma$ can be obtained from the exponential fit (see the Appendix~\ref{append:D}).
From this evidence, the genuine multipartite nonlocality from the violation of Bell inequalities $\mathcal{B}/\beta_{LR}$ exactly matches the first-order phase transition.

\section{Conclusion}
We investigate the detection of a violation of Bell inequalities with $N$-body and high-order correlations at the quantum criticality of the XXZ spin-1 chain with on-site anisotropie by employing a generalized Bell-type correlation. We apply a set of specific measurements consisting of Fourier transform states, which is considered to be optimal to detect maximally entangled states. By virtue of this optimal measurement and the symmetry of the spin-1 chain, a Bell-type correlation can be modified in a simple form. By maximizing the ratio of the Bell correlation to the classical bound $\mathcal{B}/\beta_{LR}$ we show that the quantum phase transion between the large-$D$ and AFM phases can be witnessed with $N$-body and high-order correlations. The violation of this Bell inequality in the Bell scenario $(N,3)$ is then prominently revealed in the vicinity of the first-order quantum phase transition.

Moreover, we observe a gain of entanglement as shown by an increase of bipartite entropy at the quantum phase transition. In multipartite systems there exist various type of entenglement but interestingly at the phase transition the ground state changes toward the state that maximally violate our Bell inequality which shows similarity with the generalized GHZ state.

While nonlocality is a concept that is closely related to entanglement and the presence of entanglement indicates the possibility to detect nonlocality \cite{Popescu:1992}, it needs a special care in multipartite and high dimensional composite systems. In our investigation we focus specifically on $N$-body correlations, but it is still an open question what multipartite Bell-type inequality with using $L$-body correlations (with $L\le N$) or generalized measurements \cite{Cerf:2002} would provide. Following the same formalism, this may be left as a direction of further investigations in the future.

Recently, experimental realization of the spin-1 XXZ chain with on-site anisotropy has been proposed in trapped ions \cite{Senko:2015, Cohen:2015} and implemented in ultracold atoms \cite{Chung:2021}.
This result has thus the potential to be realized in experiments beyond its theoretical demonstration.

\begin{acknowledgements}
We would like to thank to L. Amico and K. Bae for their useful comments. The DMRG method was performed by using the Tensor Network Python (TeNPy) library (version 0.7.2) \cite{tenpy}. This work was supported by National Research Foundation and Samsung Research Funding \& Incubation Center of Samsung Electronics. 
AS has been supported by a KIAS individual grant (CG070301) at Korea Institute for Advanced Study.
\end{acknowledgements}

\appendix

\section{The local realistic bound $\beta_{LR}$}
Here, we provide details for the derivation of the local realistic bound $\beta_{LR}$ in the case of the generic scenario, $N$ parties, $k$ measurement settings, and $d$ outcomes. In general, the Bell-type inequality can be constructed from the combination of joint probability distribution, which is written as
\begin{align}\label{eq:Bellineq}
\mathcal{B}=\sum_{\vec{\alpha},\vec{m}} g_{\vec{\alpha},\vec{m}}^{\vec{c}} p(\vec{\alpha}|\vec{m}) \le \beta_{LR},
\end{align}
where $g^{\vec{c}}_{\vec{\alpha},\vec{m}}$ is a Bell expression and $p(\vec{\alpha}|\vec{m})$ is the joint probability \cite{Brunner:2014}.
The vector $\vec{m}=(m_1,\cdots,m_N)$ is an array of the choice of measurements $m_l\in\{0,1,\cdots,k-1\}$ for all party and 
the vector $\vec{\alpha} = (\alpha_1(m_1),\cdots,\alpha_N(m_N))$ is an array of $N$-party measurement outcomes $\alpha_l\in\{0,1,\cdots, d-1\}$ for a given $\vec{m}$.
In the local hidden variable model, the joint probability can be expressed as the product of the local probabilities for all parties.
Then, the local realistic bound $\beta_{LR}$ in Eq.\eqref{eq:Bellineq} can be obtained by considering all deterministic values of local measurements, which is given by
\begin{align}\label{eq:def_lrb}
\beta_{LR}\equiv\max_{\vec{\alpha}} \left[ \sum_{\vec{m}} g^{\vec{c}}_{\vec{\alpha},\vec{m}} \right],
\end{align}
where the maximization is applied over $N$-party outcomes $\vec{\alpha}(\vec{m})$ for all possible choices of measurements $\vec{m}$. 
This maximization is required to consider all $d^{Nk}$ outcomes for all measurement settings.
The upper bound Eq.\eqref{eq:def_lrb} of generalized Bell correlation $\mathcal{B}$ is derived in the appendix of Ref \cite{Bae:2018}.

Generalized Bell correlation $\mathcal{B}$ Eq.\eqref{eq:lrb} is described in the form of quantum operators and the complex weight $f_n$.
Since the $N$-body correlation, the expectation value of the quantum operators, can be written in terms of the probabilities,
a Bell expression $g^{\vec{c}}_{\vec{\alpha},\vec{m}}$ can be derived from generalized Bell correlation Eq.\eqref{eq:lrb}.
The main idea of this derivation is that one should take into account the deterministic values of the measurement in Eq.\eqref{eq:lrb}.
Then, the Bell expression  $g^{\vec{c}}_{\vec{\alpha},\vec{m}}$ in terms of the weight $f_n$ is given by
\begin{equation}\label{eq:gfrelation} 
g^{\vec{c}}_{\vec{\alpha},\vec{m}}=2\,\text{Re}\left[\sum^{d-1}_{n=1}f_n\omega^{n[\vec{c}\cdot(\vec{\alpha}+\frac{\vec{m}}{k})]}\right],
\end{equation}
where $\omega=\exp(2\pi i/d)$, the vector $\vec{c}=(c_1,\cdots,c_N)$ with $c_l$ being either +1 or -1 for any party $l$, and $\text{Re}\,z$ stands for the real part of the complex number z. 
By using Eq.\eqref{eq:gfrelation}, the sum of the Bell expression $g^{\vec{c}}_{\vec{\alpha},\vec{m}}$ over $\vec{m}$  which is denoted by $\mathcal{S}(\vec{\alpha})$ can be simplified as
\begin{align}\nonumber
\mathcal{S}(\vec{\alpha})&\equiv \sum_{\vec{m}} g^{\vec{c}}_{\vec{\alpha},\vec{m}} =2 \sum_{\vec{m}} \text{Re} \left[ \sum_{n=1}^{d-1} f_n\omega^{n[\vec{c}\cdot(\vec{\alpha}+\frac{\vec{m}}{k})]} \right] \\
\label{eq:sum_g}&=\sum_{\vec{m}} \sum_{n=1}^{d-1} 2|f_n| \cos\left[\theta_n + \frac{2\pi n }{d}\vec{c}\cdot\left(\vec{\alpha}+\frac{\vec{m}}{k}\right)\right],
\end{align}
where $f_n=|f_n|e^{i\theta_n}$ with $0\le \theta_n\le 2\pi$.
Therefore, one can express the local realistic bound that corresponds to Eq.\eqref{eq:lrb} by applying $k=2$ and $d=3$ in Eq.\eqref{eq:sum_g}.
It is noted that the computational difficulty of maximization arises as the values of $N$, $k$, and $d$ increase.
Detail derivations of Eqs.\eqref{eq:def_lrb}, \eqref{eq:gfrelation}, and \eqref{eq:sum_g} are well explained in Ref \cite{Bae:2018}.

\begingroup 
\begin{table}[b]
\begin{tabular}{c | c | c | c | c}
\hline\hline
$\alpha_1(0)$ & $\alpha_1(1)$ & $\alpha_2(0)$ & $\alpha_2(1)$ & Eq.\eqref{eq:sum_g_chsh}\\
\hline
0  &  0 &  0 & 0  & 2 \\
0  &  0 &  0 & 1  & 2 \\
0  &  0 &  1 & 0  & -2 \\
0  &  0 &  1 & 1  & -2 \\
0  &  1 &  0 & 0  & 2 \\
0  &  1 &  0 & 1  & -2 \\
0  &  1 &  1 & 0  & 2 \\
0  &  1 &  1 & 1  & -2 \\
1  &  0 &  0 & 0  & -2 \\
1  &  0 &  0 & 1  & 2 \\
1  &  0 &  1 & 0  & -2 \\
1  &  0 &  1 & 1  & 2 \\
1  &  1 &  0 & 0  & -2 \\
1  &  1 &  0 & 1  & -2 \\
1  &  1 &  1 & 0  & 2 \\
1  &  1 &  1 & 1  & 2 \\
\hline\hline
\end{tabular}
\caption{16 possible measurement outcomes for all choices of measurements $\alpha_1(0), \alpha_1(1), \alpha_2(0)$, and $\alpha_2(1)$ in two-party two-outcome systems with two measurement settings.}
\label{table}
\end{table} 
\endgroup

\section{Computing local realistic bound: an example for CHSH and CGLMP inequalities} \label{append:B}
When $N\le 10$, $k=2$ and $d=3$ that we deal with, the calculation of the local realistic bound is rather complicated.
For simplicity, let us give first an example to calculate the local realistic bound of the Clauser-Horne-Shimony-Holt (CHSH) inequality \cite{Clauser:1969}. 
To discuss this Bell inequality, one considers the scenario that each of two parties performs two different dichotomic measurements, i.e., $N=2, k=2$, and $d=2$.
In this scenario, an Bell expression for the CHSH inequality can then be written as
\begin{align}
g^{CHSH}_{\vec{\alpha},\vec{m}} &=  (-1)^{\alpha_1(m_1)+\alpha_2(m_2)+m_1m_2},
\end{align}
where the vector $\vec{c}=(1,1)$ is chosen \cite{Bae:2018}. 
The sum of $g^{CHSH}_{\vec{\alpha},\vec{m}}$ over the vector $\vec{m}$ is then written as
\begin{align}\nonumber
\sum_{m_1,m_2=0}^1 g^{CHSH}_{\vec{\alpha},\vec{m}} & = (-1)^{\alpha_1(0)+\alpha_2(0)} + (-1)^{\alpha_1(0)+\alpha_2(1)}\\
\label{eq:sum_g_chsh} & + (-1)^{\alpha_1(1)+\alpha_2(0)} - (-1)^{\alpha_1(1)+\alpha_2(1)}.
\end{align}
In order to maximize Eq.\eqref{eq:sum_g_chsh}, one should consider all $d^{Nk}=2^4$ outcomes for all measurement choices, which is organized in Table I.  
Therefore, one can find that the local realistic bound of the CHSH inequality is $\beta_{LR}^{CHSH} = \max_{\alpha_1,\alpha_2} \sum_{\vec{m}}g^{CHSH}_{\vec{\alpha},\vec{m}}=2$.

In this way, the local realistic bound can be found for more generic cases.
Since generalized Bell correlation is applied to the spin-1 model Eq.\eqref{eq:spin1H},
we particularly choose two choices of measurements with three outcomes, i.e., $k=2$ and $d=3$. 
A function $\mathcal{S}(\vec{\alpha})$ from Eq.\eqref{eq:sum_g} can then be written as

\begin{align}\label{eq:sum_g_n23}
\mathcal{S}(\vec{\alpha}) =\sum_{n=1}^2 \left[2|f_n|\sum^1_{\{m_l\}=0}\cos\left(\theta_n + \frac{2\pi n}{3}S_{\alpha} + \frac{\pi n}{3} S_{m}\right) \right], 
\end{align}
where $S_{\alpha}=\sum_{l=1}^N c_l\alpha_l(m_l)$ and $S_{m}=\sum_{l=1}^N c_lm_l$ with choosing $c_l=(-1)^{l-1}$ for $1\le l\le N$.
As the number of party $N$ increase, an exponential amount of cosine terms emerge in Eq.\eqref{eq:sum_g_n23}. 
Furthermore, a $d^{kN}$ number of measurement outcomes should be considered to evaluate the local realistic bound $\max_{\vec{\alpha}}\mathcal{S}(\vec{\alpha})$. 
Due to this computational complexity, we deal with generalized Bell inequalities up to $N=10$ in this work.

Here, we show an example about the local realistic bound of the Collins-Gisin-Linden-Masser-Popescu (CGLMP) inequality  \cite{Collins:2002}. 
Eq.\eqref{eq:sum_g_n23} for the case of $N=2, k=2$, and $d=3$ is described as

\begin{align}\nonumber
\mathcal{S}(\vec{\alpha})&=\sum^2_{n=1} 2|f_n| \left\{ \cos\left(\theta_n + \frac{2\pi n}{3}[\alpha_1(0)-\alpha_2(0)] \right) \right. \\
\nonumber &\;\;\;\;\;\;\;\;\;\;+ \left.\cos\left(\theta_n + \frac{2\pi n}{3}[\alpha_1(0)-\alpha_2(1)] - \frac{\pi n}{3}\right) \right. \\
\nonumber &\;\;\;\;\;\;\;\;\;\; \left. + \cos\left(\theta_n + \frac{2\pi n}{3}[\alpha_1(1)-\alpha_2(0)] + \frac{\pi n}{3}\right) \right. \\
\label{eq:sum_g_223} &\;\;\;\;\;\;\;\;\;\;\;\;\;\; \left.+ \cos\left(\theta_n + \frac{2\pi n}{3}[\alpha_1(1)-\alpha_2(1)] \right) \right\},
\end{align}
where the weights $f_n$ for the CGLMP inequality are given by $f^{CGLMP}_1=\omega^{1/4}/2\sqrt{3}$ and $f^{CGLMP}_2=\omega^{1/2}/2$ for $d=3$ \cite{Bae:2018}.
There are $3^{4}$ possible outcomes for all choices of measurements $\alpha_1(0)$, $\alpha_1(1)$, $\alpha_2(0)$, and $\alpha_2(1)$ similar to Table~\ref{table}.
So if we calculate $\mathcal{S}$ of Eq.\eqref{eq:sum_g_223} for the CGLMP case,
one can then find $\mathcal{S}(\vec{\alpha})\in\{-4,1,2\}$ for all possible outcomes $\vec{\alpha}$ and the local realistic bound of the CGLMP inequality becomes $\beta_{LR}^{CGLMP}=\max_{\vec{\alpha}}\mathcal{S}(\vec{\alpha})=2$.

\section{Maximization of $\mathcal{B}/\beta_{LR}$ over $f_1$ and $f_2$} \label{append:C}
From Eq.\eqref{eq:n23Bellcorr}, Bell correlation $\mathcal{B}$ for a ground state is described as a real-value function of variables $f_1$, $f_2$, and $\theta_\nu$.
The local realistic bound $\beta_{LR}$ also can be considered as a function of $f_1, f_2$.
By using Eq.\eqref{eq:n2dBellop} and Eq.\eqref{eq:lrb}, the ratio of Bell correlation to the local realistic bound can be written as a real-value function of $\tilde{f}, \theta_1, \theta_2$, and $\theta_\nu$,

\begin{widetext}
\begin{align}\nonumber
\mathcal{F}(\tilde{f},\theta_1,\theta_2,\theta_\nu) &\equiv \frac{\mathcal{B}(f_1,f_2,\theta_\nu)}{\beta_{LR}(f_1,f_2)} \\
\label{eq:belltolrb}&= 2^{N}\frac{\cos(\theta_1-\theta_\nu)\langle \hat{J}_1\hat{J}^{-1}_2\cdots \hat{J}^{-1}_N \rangle + \tilde{f}\cos(\theta_2-2\theta_\nu)\langle \hat{J}_1^2\hat{J}^{-2}_2\cdots \hat{J}^{-2}_N \rangle}{\max_{\{\alpha_l\}_{l\le N}} \left[\sum_{\{m_l\}_{l\le N}} \cos \left(\theta_1 + \frac{2\pi}{3}\vec{c}\cdot\vec{\alpha}+ \frac{\pi }{3}\vec{c}\cdot\vec{m}\right) + \tilde{f}\cos \left(\theta_1 + \frac{2\pi}{3}\vec{c}\cdot\vec{\alpha}+ \frac{\pi }{3}\vec{c}\cdot\vec{m}\right) \right]},
\end{align}
\end{widetext}
where 
$f_n = |f_n|e^{i\theta_n}$ and $\tilde{f}\equiv|f_2|/|f_1|$. 
The values of first-order and second-order $N$-body correlations in Eq.\eqref{eq:belltolrb} are determined by the ground states depending on $J_z$ and $D$.
Here, we obtain $N$-body correlations with $N\le8$ for the ground states obtained from the exact diagonalization method.
For $N=10$, we evaluate correlations based on the tensor network representation and the ground states are obtained from the DMRG method.

In order to find the maximum $\mathcal{F}(\tilde{f},\theta_1,\theta_2,\theta_\nu)$ Eq.\eqref{eq:belltolrb} for a given ground state,
we exploit the brute force method, one of the global optimization tools in the scipy library (scipy.optimize.brute) \cite{2020SciPy-NMeth}.
This numerical evaluation results in finding the maximal values of $\mathcal{F}$ and corresponding variables $\tilde{f}_{\max}, \theta_{1,\max}, \theta_{2,\max}$, and $\theta_{\nu,\max}$. 
These specific data are described in Sec.\ref{sec:qv} as tuning the parameters $J_z$ and $D$.

\section{Scaling Behaviors of Bell correlations divided by the local realistic bound} \label{append:D}

\begin{figure}[b!]
    \centering
    (a) $J_z=4$
    \includegraphics[width=0.47\textwidth]{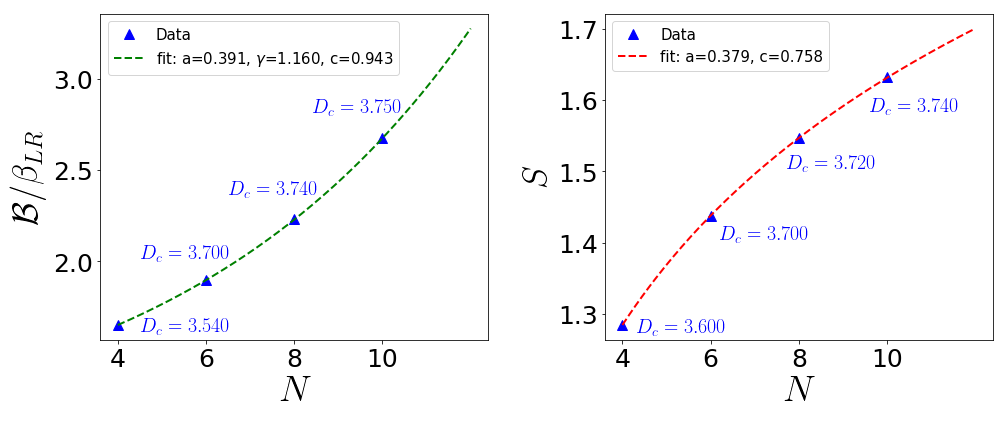}
    (b) $J_z=5$
    \includegraphics[width=0.47\textwidth]{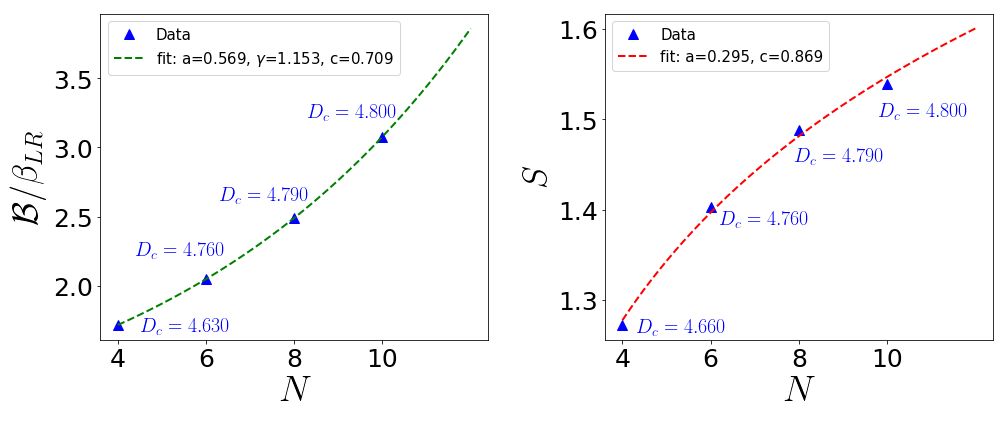}
    (c) $J_z=6$
    \includegraphics[width=0.47\textwidth]{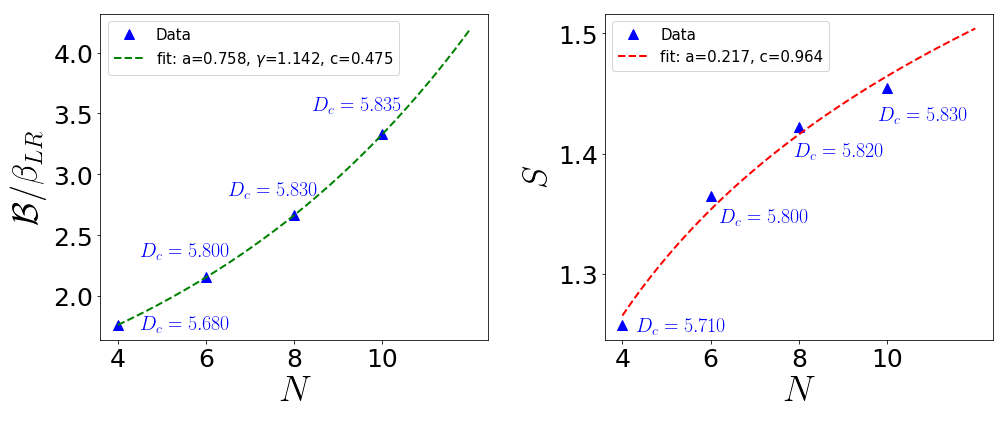}
    \caption{Scaling behaviors of the ratio of Bell correlation to the local realistic bound $\mathcal{B}/\beta_{LR}$ (left) and the entanglement entropy $S$ (right) at the quantum criticality. 
    The green line is the exponential fit $\mathcal{B}/\beta_{LR} = a \gamma^N+c$ and the red line is the logarithmic fit $S = a\log N + c$.
    }
    \label{fig:scaling}
\end{figure}

At this criticality, Bell correlation divided by the local realistic bound $\mathcal{B}/\beta_{LR}$ tends to increase as the number of particles $N$ increases.
At the first-order phase transition between the large-$D$ and the antiferromagnetic phases, we expect that $\mathcal{B}/\beta_{LR}$ with $\tilde{f}^v_{\max}$ behaves as an exponential function of $N$, i.e., $\mathcal{B}/\beta_{LR}\sim \gamma^N$ with a real $\gamma$, which is plotted on Fig.~\ref{fig:scaling}. 
For the exponential fit $a \gamma^N +c$, the variances of $a$, $\gamma$, and $c$ are smaller than $10^{-4}$ for $J_z=4$, $10^{-5}$ for $J_z=5$, and $10^{-6}$ for $J_z=6$.

In order to be compared to the many-body entanglement and understand the many-body features, let us introduce the entanglement entropy which tells an amount of entanglement for bipartite pure states.
For a given pure state $|\psi\rangle_{AB}$ on the composite system $AB$, the mathematical definition of entanglement entropy is given by
\begin{align}
S \equiv \text{Tr}[\rho_A\log\rho_A] = \text{Tr}[\rho_B\log\rho_B],
\end{align}
where $\rho_A$ (or $\rho_B$) stands for the reduced density matrix of the state $|\psi\rangle$ on the subsystem $A$ (or $B$). 
It is known that the in the one-dimensional systems the entanglement entropy for large $N$ converges the constant in the gapped systems and diverges logarithmically in the gapless ones \cite{Vidal:2003}. 
This feature also holds for quantum criticalities in the 1D spin-1 XXZ model with on-site anisotropy.
In this work, the reduced density matrices $\hat{\rho}_{N/2}$ for the subsystem of size $N/2$ are chosen to investigate the entanglement.
At the criticality between the large-$D$ and the antiferromagnetic phases, the entanglement entropy for even $N\le 10$ is plotted in Fig.~\ref{fig:scaling}. For small $N$, the fitting curve behaves as $S(N)\sim \log N$ but its accuracy does not exact.
As the particle number $N$ grows, the entanglement entropy is expected to attain a saturation value.

\end{document}